\documentclass[twocolumn,showpacs,prl]{revtex4}
\usepackage{graphicx}

\begin{document}

\title{Universal Out-of-Equilibrium Transport in Kondo-Correlated Quantum
Dots}
\author{A. A. Aligia}
\affiliation{Centro At\'{o}mico Bariloche and Instituto Balseiro, Comisi\'{o}n Nacional
de Energ\'{\i}a At\'{o}mica, 8400 Bariloche, Argentina}
\date{\today}
\maketitle

Two recent works attempt to extend results for the conductance $G$ through a
quantum dot described by the particle-hole symmetric (PHS) impurity Anderson
model (IAM) out of the PHS case using renormalized perturbation theory in $U$
(RPTU) up to $U^{2}$ \cite{sca,mu}. Contrary to what is stated in the 
\textit{note added} of Ref. \onlinecite{mu}, previous results for the PHS
case (where the occupation of the dot is $n=1$ by symmetry) and general
coupling to the left and right leads $\Gamma _{L(R)}$ and chemical
potentials $\mu _{L(R)}=\pm \alpha _{L(R)}eV$ \cite{sela}, are recovered by
the first approach [Eq. (30) of Ref. \onlinecite{sca}]. The dependence on
temperature $T$ is also correct. 
Ward identities are trivially satisfied because they were used [see the paragraph above Eq. (23)].
A limitation of this approach is that out
of the PHS case, the coefficients of the expansion of $G$ in terms of $T$
and voltage $V$ contain derivatives of $n$ or the real part of the retarded
self energy $\Sigma ^{r}$, which for an interacting system seem to depend on
high energy properties hard to capture in a Fermi liquid approach.
Exceptions are the linear term in $T$ for $V=0$ (which vanishes) and the
linear term in $V$ for $T=0$ (addressed below).

Instead, Ref. \onlinecite{mu} provides explicit expressions for all
coefficients up to second order in $T$ and $V$ for $\alpha _{L}\Gamma
_{R}=\alpha _{R}\Gamma _{L}$ and $n\rightarrow 1$. Unfortunately, the
authors have made mistakes in the evaluation of the lesser quantities $%
\Sigma ^{-+}$ and $G^{-+}$ (-$\Sigma ^{<}$ and $G^{<}$ in our notation)
already in the PHS case. This implies that also the greater quantities are
incorrect. While using conservation of the current (CC), lesser and greater
functions can be eliminated from the expression of $G$, they play a crucial
role precisely in this conservation [Eqs. (80) to (89) of Ref. %
\onlinecite{sup}], and therefore the approach seems unreliable. One
incorrect result is that $\Sigma _{MBK}^{<}(\omega )=2if_{eff}(\omega )$Im$%
[\Sigma ^{r}(\omega )]$, where $f_{eff}(\omega )$ is the average of the Fermi
function at the two leads, weighted by the corresponding $\Gamma _{\nu }$.
In addition, the authors claim to demonstrate that the term proportional to
the non-interacting lesser Green function $g^{<}$ in the expression for $%
G^{<}$ [first term in Eq. (73) of Ref. \onlinecite{sup}] vanishes (although
it can be written as $2if_{eff}(\omega )\Delta |G^{r}|^{2}$  
[Eqs. (7), (8) of Ref. \onlinecite{none}]), and uses
this result to claim that $G_{MBK}^{<}=-|G^{r}|^{2}\Sigma _{MBK}^{<}$.
CC would follow from the form of $G_{MBK}^{<}$, $\Sigma _{MBK}^{<}$, and 
known relations between the different Green functions.

Unfortunately, the demonstration is flawed because Eq. (76) of Ref. 
\onlinecite{sup} is used, which misses the term $2i\Delta $. The correct
form of this equation is $(G^{r})^{-1}-(G^{a})^{-1}=\Sigma ^{a}-\Sigma
^{r}+2i\Delta $.
This comes trivially from the definition of 
$G^r$ [third line below Eq. (13) of Ref. 2] and its complex conjugate $G^a$.

\begin{figure}[!ht]
\includegraphics[width=.90\linewidth]{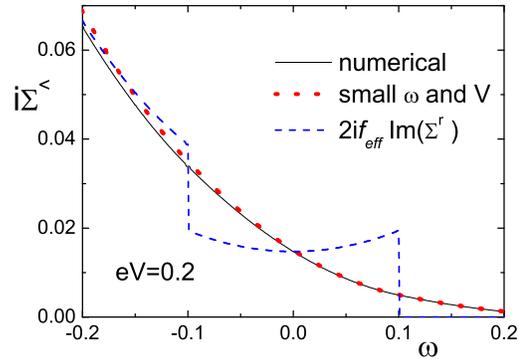}
\caption{(Color online) Lesser self energy as a function of frequency of the
PHS IAM for renormalized interaction $\tilde{u}=\widetilde{U}/(\protect\pi 
\widetilde{\Delta }) = 1$, $\Gamma_R=\Gamma_L$, $\protect\alpha_L=\protect%
\alpha_R$, $T=0$, and $eV=0.2 \widetilde{\Delta }$. $\widetilde{\Delta }$ is
taken as the unit of energy.}
\label{sigl}
\end{figure}

In Fig. \ref{sigl} we compare $\Sigma _{MBK}^{<}(\omega )$ at $T=0$ with the
correct one, obtained integrating numerically the RPTU expressions \cite%
{none,her}. We also display the analytical result \cite{sca,her} up to
total second order in $\omega $ and $V$ [Eq. (20) or Ref. \onlinecite{sca}].
As it is known \cite{her}, the correct result is continuous. Instead, $%
\Sigma _{MBK}^{<}$ has jumps at $\mu _{L}$ and $\mu _{R}$, and strongly
disagrees with the correct result except at energies far away from both 
$\mu_{\nu }$. 

It is difficult to say how these mistakes affect the reported expansion
coefficients. The linear term in $V$ can be written in the form $%
c_{VE_d}=2(\alpha_L-\alpha_R)\cos (\pi n/2)$, which coincides with the
result of Ref. \onlinecite{sca} taking $\alpha_L \Gamma_R = \alpha_R
\Gamma_L $ and $n \rightarrow 1$. In any case, CC
is an essential requisite for any nonequilibrium theory. RPTU conserves the
current in the PHS case and up to order $V^3$ in the general case for $T=0$ 
\cite{sca}.

This work was sponsored by PIP 1821 of CONICET and PICT R1776 of the ANPCyT,
Argentina. A.A.A. is partially supported by CONICET.

\end{document}